\documentclass[english,aps,pre,showpacs,superscriptaddress,amsmath,amssymb,twocolumn]{revtex4}
\usepackage[T1]{fontenc}
\usepackage[latin9]{inputenc}
\usepackage{textcomp}
\usepackage{amstext}
\usepackage{graphicx}

\makeatletter

\newcommand{\lyxmathsym}[1]{\ifmmode\begingroup\def\b@ld{bold}
  \text{\ifx\math@version\b@ld\bfseries\fi#1}\endgroup\else#1\fi}

\@ifundefined{textcolor}{}
{%
 \definecolor{BLACK}{gray}{0}
 \definecolor{WHITE}{gray}{1}
 \definecolor{RED}{rgb}{1,0,0}
 \definecolor{GREEN}{rgb}{0,1,0}
 \definecolor{BLUE}{rgb}{0,0,1}
 \definecolor{CYAN}{cmyk}{1,0,0,0}
 \definecolor{MAGENTA}{cmyk}{0,1,0,0}
 \definecolor{YELLOW}{cmyk}{0,0,1,0}
 }

\usepackage{epsfig}
\usepackage{bm}
\usepackage{bbm}

\usepackage{epic}
\usepackage{eepic}
\usepackage{pifont}
\@ifundefined{definecolor}
 {\usepackage{color}}{}

\usepackage{nicefrac}
\input epsf
\input rotate

\newcommand{\be}{\begin{equation}}
\newcommand{\ee}{\end{equation}}
\newcommand{\bea}{\begin{eqnarray}}
\newcommand{\eea}{\end{eqnarray}}

\newcommand{\comment}[1]{}

\makeatother

\usepackage{babel}

\begin{document}

\title{New analytical progress in the theory of vesicles under linear flow}

\author{Alexander Farutin}

\affiliation{Laboratoire de Spectrom\'etrie Physique, UMR5588, 140 avenue de la physique,
Universit\'e Joseph Fourier Grenoble, and CNRS, 38402 Saint Martin d'H\`eres,
France }
\author{Thierry Biben}

\affiliation{ Universit\'e de Lyon, F-69000, France; Univ. Lyon 1, Laboratoire PMCN; CNRS, UMR 5586; F-69622 Villeurbanne Cedex}

\author {Chaouqi Misbah $^1$}

\begin{abstract}
Vesicles are becoming a quite popular model for the study of red blood cells (RBCs). This is a free boundary problem which is rather difficult to handle theoretically. Quantitative computational approaches constitute  also a challenge. In addition, with numerical studies, it is not easy to scan within a reasonable time the whole
parameter space. Therefore, having quantitative analytical results is an essential advance that provides deeper understanding of observed features and can be used to accompany and possibly guide further numerical development.
In this paper
shape evolution equations for a vesicle in a shear flow are derived analytically with precision being cubic (which is quadratic in previous theories) with regard to  the deformation of the vesicle relative to  a spherical
shape.
The phase diagram distinguishing regions of parameters where different types of motion (tank-treading, tumbling and vacillating-breathing)
 are manifested is presented. This theory reveals unsuspected  features:  including higher order terms and harmonics (even if they are not directly excited by the shear flow) is necessary, whatever the shape is close to a sphere. Not only does this theory cure a quite large quantitative discrepancy between previous theories
 and recent experiments and numerical studies, but also it reveals a new phenomenon: the VB mode band in parameter space, which is believed to saturate after a moderate shear rate, exhibits a striking widening beyond a critical shear rate. The widening results from excitation of fourth order harmonic.
The obtained phase diagram is in a remarkably good agreement with
recent  three dimensional numerical simulations based on the
boundary integral formulation. Comparison of our results with experiments is systematically made.
\end{abstract}


\maketitle

\section{Introduction}
A vesicle is a closed simply connected membrane separating two liquids. Its dynamics in an ambient flow is an important research target due to various medical applications. For instance,
they are employed as biochemical reactors \cite{Noireaux-Libchaber2004,Karlsson2004}, as vectors for targeted
drug and gene delivery \cite{Allen2004,Gregoriadis1995}, and as artificial cells for hemoglobin
encapsulation and oxygen transport \cite{Arifin2005}.

One can also regard a vesicle, from mechanical point of view, as a simplified model of a living
cell. The most prominent biological counterpart for a vesicle is the red blood cell (RBC).
The vesicle  system allows to identify the elementary processes of
visco-elastic properties of  cells moving passively in a fluid. The
study of a single vesicle under a shear flow is one of the simplest
nonequilibrium example. Nevertheless,  this problem gives rise to a very complicated
dynamics with different types of vesicle motion because of the
nontrivial competition between the interactions of straining and rotational
parts of shear flow with the vesicle. Three main types of motion for
a vesicle in a shear flow have been identified so far: (i)
tank-treading (TT), when shape and orientation of the vesicle is a
steady-state, (ii) tumbling (TB), when the vesicle makes full
rotations, and (iii) vacillating-breathing (VB, aka trembling or
swinging), when the longest axis of the vesicle oscillates around
certain direction with the shape strongly changing during these
oscillation cycles. Since many properties of a vesicle (e.g. the
effective viscosity) strongly depend on its type of motion, it is
important to define the phase diagram of a vesicle in a shear flow,
predicting which type of motion will be realized for each set of
parameters defining the vesicle dynamics.

A widely accepted model for a vesicle proposes the following
assumptions: zero Reynolds number limit, membrane impermeability and
local inextensibility, and the continuity of velocity field and mechanical stress across
the membrane. The force exerted on the membrane by the liquids is
balanced by the membrane rigidity force that is calculated from the
energy of membrane bending taken as\cite{H73}
\begin{equation}\label{HE}E=\frac{\kappa}{2}\int(2H)^2dA+\int
ZdA,\end{equation} where $\kappa$ is the bending rigidity
coefficient of the membrane, $H$ is the mean curvature and $Z$ is a
Lagrange multiplier, ensuring the local membrane inextensibility.
Under these assumptions, and after a certain choice of units for all
variables, the vesicle dynamics can be completely described by three
dimensionless numbers (see for example \cite{Kaoui2009}): the
viscosity contrast
\begin{equation} \lambda={\eta_{int}\over \eta_{ext}},\end{equation}
the excess area relative to a sphere \begin{equation}
\Delta=Ar_0^{-2}-4\pi,\end{equation} and the capillary number
\begin{equation} C_a={\eta_{ext}\dot{\gamma}r_0^3\over
\kappa}.\end{equation} Here $\eta_{int,ext}$ are the viscosities of
fluids inside and outside the membrane respectively, $A$ is the
surface area of the vesicle, $r_0=\left ({3V\over 4\pi} \right
)^{1/3}$ is the radius of a sphere containing the same volume as the
vesicle, and $\dot{\gamma}$ is the shear rate. It is convenient to
choose $r_0$ as the unit of distance so that the volume of the
vesicle is \begin{equation}\label{vc}V={4\pi\over 3}.\end{equation}
We also use $\eta_{ext}$ as the unit of viscosity and
$\dot{\gamma}^{-1}$ as the unit of time.

Several studies have been recently  dedicated to the determination
of the phase diagram regarding different types of motion  of an
almost spherical vesicle in a shear flow. These studies were
 theoretical\cite{M06,G07,D07,L07}, experimental \cite{Ma06,K06,De09,De092},
and  numerical \cite{Noguchi2007,B09}. Those  theoretical works
treat the membrane deviation from a spherical shape as a
perturbation ($\epsilon=\sqrt{\Delta}$ is the small expansion parameter), represented as a series of spherical harmonics.
 Then it is suggested that one or two first orders in the expansion { of the shape evolution equations} must be retained. Generally, the resulting equations differ only because the authors propose distinct rules of neglection. The Leading Order Theory\cite{M06} (LOT) keeps only the terms of order of the imposed flow and the effects of the vorticity in the final equations. This theory is precise up to $O(\epsilon)$ in the expansion scheme. Later studies  have included higher order terms in the expansion (up to $O(\epsilon^2))$. Lebedev et al. \cite{L07} included the next   order terms in the membrane Helfrich force (as compared to LOT), while another study (hereafter called
 Higher Order Theory\cite{D07} (HOT)) accounted for  the next order terms not only in the Helfrich force but also in the hydrodynamic field as well.

  In the theories of Ref. \cite{D07} and Ref. \cite{L07}, as well as
  in numerical studies based on dissipative particles dynamics \cite{Noguchi2007},  a saturation of the VB/TB phase border
  for large enough $C_a$ was suggested-- i.e. the value of $\lambda$ at which the transition from VB to tumbling occurs
  becomes almost independent on the capillary number (or equivalently on shear rate).

Recent advances in three dimensional numerical computations based on
the boundary integral formulation \cite{B09} have made it possible
to study vesicle dynamics quantitatively. The results of this study
show, that the loss of stability of the TT solution occurs at values
of $\lambda$ significantly higher than those predicted by analytical
calculations\cite{D07,L07}. Furthermore, no saturation of
the VB/TB phase border upon increasing shear rate (or $C_a$) was found. These
results provide a new input that is worth understanding from the
analytical point view.

The main objective of this work is to investigate the reasons for the discrepancies between the results
 provided by the recent numerical simulations \cite{B09} and analytical theories.
 It was pointed out  in Ref.\cite{L07} that the critical values of $\lambda$ at which the steady-state solution loses its stability, scales as $O(\Delta^{-1/2})$ for a fixed shear rate. However, this  behaviour, which also agrees with
 the result of Keller and Skalak \cite{K82}, does not provide the correct information about  the next order terms, as will be seen here. The next correction turns out to cause a rather big shift of the phase borders.

 This peculiar behavior (i.e. that the critical $\lambda$ diverges at vanishing
 $\Delta$) confers to the vesicle problem a special status:
 in the small $\Delta$ limit, {\it one further order and only one}
 in  the expansion scheme in powers of $\Delta$ is needed as compared to previous studies.
 As a consequence, it will be shown, in particular, that the next order term, previously unaccounted for, survives whatever small the deviation
 from a sphere is. This is the source of deviation between the recent numerical work and the existing analytical
  theories.
 As will be seen inclusion of one more order in the expansion will  allow us to extract a phase diagram that
 is in quantitative agreement with the results of numerical simulations\cite{B09}.
 It will also be shown  that  fourth order spherical harmonics (in previous theories only second order harmonics were included) can not be neglected, and especially beyond a certain shear rate. Indeed, it will be revealed that
 these harmonics give rise to a new important feature recently revealed  in numerical
 simulations \cite{B09}. More precisely, the VB/TB phase border does
 not saturate for fixed $\Delta$ when shear rate exceeds  a certain value, rather a significant
 widening is revealed, in contrast to previous theoretical \cite{L07,D07}, numerical \cite{Noguchi2007}
 and experimental investigations \cite{De09}.

\section{Shape evolution equations}

\subsection{Basic definitions and the expansion scheme}

 Here we perform
 the expansion of the shape evolution equations one order higher as compared to the most advanced of previous papers\cite{D07}.  Since a slightly deflated vesicle in the absence of flow takes a centro-symmetric shape, and the shear flow has a symmetry center in every point, we consider only symmetric shapes.

  Naturally, we take  the origin of coordinates at  the center of the vesicle moving with the velocity of the undisturbed shear flow. That the vesicle moves with the same  velocity as  the applied shear flow is a consequence of the  symmetry of the problem. We use the conventional parametrization of the vesicle membrane with reference  points belonging to a sphere of radius  unity  \begin{equation}\boldsymbol{R}=\boldsymbol{x}[1+f(\boldsymbol{x})].\end{equation}The shape function $f(\boldsymbol{x})$ is then expanded in spherical harmonics of $\boldsymbol{x}.$ Note that the shape depends also on time, but the temporal variable will be specified only when need be.
  Following Ref.\cite{G07,D07} we introduce a formal expansion parameter $\varepsilon=\sqrt{\Delta}$ which helps classifying different orders. Only second order harmonics are present to the order of $O(\varepsilon)$ in the equilibrium shape function of an almost spherical vesicle in the absence of flow. They are also the only harmonics that can interact with linear flow directly. For consistency considerations, pushing the expansion to next order implies that  the fourth and zeroth order harmonics  cannot be neglected. They enter dynamics as a result
  of interactions between the vesicle shape and the Helfrich force and the  flow; they are of order $O(\varepsilon^2).$ The shape function does not contain spherical harmonics of odd orders owing to the centro-symmetry. The spherical harmonics of even orders higher than $4$ have an amplitude  $O(\varepsilon^3)$ and thus are neglected. As follows from the inequality $$1=\frac{3V}{4\pi}=\int R(\boldsymbol{x})^3\frac{d^2x}{4\pi}\ge\left[\int R(\boldsymbol{x})\frac{d^2x}{4\pi}\right]^3=R_0^3$$ the average radius of the vesicle $R_0$ is not exactly 1. Therefore   a negative correction must  be added to it in order  to satisfy (\ref{vc}). This correction  is of order $O(\varepsilon^2)$ and
  leads to a non-zero coefficient for zero-order spherical harmonic in the expansion of
  $f(\boldsymbol{x}),$ which will be denoted $f_0$ below.

The  expansion takes formally the form:
\begin{equation}\label{f}f(\boldsymbol{x})=\varepsilon^2f_0+\varepsilon f_2+\varepsilon^2 f_4+O(\varepsilon^3),
\end{equation} $$f_k=\sum^k_{l=-k}f_{k,l}(t) Y_{k,l}(\boldsymbol{x}).$$

 There are various definitions of spherical harmonics, here we use those that satisfy the following normalization conditions:
$$\int{Y_{k,l}(\boldsymbol{x})Y^*_{k,l}(\boldsymbol{x})d^2x=\frac{4\pi(k!)^2}{(2k+1)(k+l)!(k-l)!}}$$ However, all equations are written in  universal form and are valid for any rescaling of spherical harmonics.

 The condition (\ref{vc}) together with  the definition $\Delta=\varepsilon^2$ provide two additional constraints on the coefficients of the expansion (\ref{f}). We use the former to express $f_0$ through $f_{2,l}$
\begin{equation}\label{f0}\int f_0(\boldsymbol{x})d^2x=-\int f_2(\boldsymbol{x})^2d^2x-\frac{\varepsilon}{3}\int f_2(\boldsymbol{x})^3d^2x+O(\varepsilon^2)\end{equation} and the latter to norm the coefficients $f_{2,l}$ \begin{equation}\label{ea}2\int
f_2(\boldsymbol{x})^2d^2x-\frac{2\varepsilon}{3}\int
f_2(\boldsymbol{x})^3d^2x=1+O(\varepsilon^2).\end{equation}  We
retain one more order in these expressions compared to previous
works for the sake of consistency. Note that the volume and the
surface have  been expanded to the order $O(\epsilon^3)$ in order
to obtain the equalities (\ref{f0}) and (\ref{ea}).

\subsection{Derivation of the evolution equations of the $f_k$ from
the boundary integral formulation}

We shall adopt here a different spirit from that of  previous
theories \cite{M06,D07,L07,G07}. There  the Lamb solution is used (a
solution of the Stokes equations inside and outside the vesicle). We
find it convenient to take the boundary integral formulation as a
starting point. The present  spirit is equivalent to using the Lamb
solution, but it has  the advantage that
the boundary conditions at the membrane are already implemented in the boundary integral formulation.
Indeed,
the Stokes equations together with the boundary conditions on the
membrane, and far away from it, can be converted into
 boundary integral formulation \cite{Pozrikidis_book}. This leads to the
 following integral equation \begin{equation}\label{ie}\begin{split}&v_l(\boldsymbol{x})(1+\lambda)=
 \\&=2u_l(\boldsymbol{x})+2\int G_{jl}\left(\boldsymbol{R}(\boldsymbol{x})-\boldsymbol{R}(\boldsymbol{x}')\right)
 F_j(\boldsymbol{x}')d^2R(\boldsymbol{x}')+\\&+(1-\lambda)\int K_{jlm}\left(\boldsymbol{R}(\boldsymbol{x})-
 \boldsymbol{R}(\boldsymbol{x}')\right)v_j(\boldsymbol{x}')N_m(\boldsymbol{x}')d^2R(\boldsymbol{x}').\end{split}
 \end{equation} Here $\boldsymbol{v}(\boldsymbol{x})$ is the actual velocity
 of point $\boldsymbol{R}(\boldsymbol{x}),$ $\boldsymbol{u}(\boldsymbol{x})$
 is the velocity of point $\boldsymbol{R}(\boldsymbol{x})$ in the imposed flow,
  $\boldsymbol{F}(\boldsymbol{x})$ is the membrane rigidity force at point $\boldsymbol{R}(\boldsymbol{x}),$
  $\boldsymbol{N}(\boldsymbol{x})$ is the outward pointing normal to the vesicle surface at
   point $\boldsymbol{R}(\boldsymbol{x}).$ The integration is taken over the surface of the vesicle
   and $d^2R(\boldsymbol{x}')$ is the surface area element. The integration kernels have the following
   form: $$G_{ij}(\boldsymbol{R})=\frac{1}{8\pi}
   \left(\frac{\delta_{ij}}{R}+\frac{R_iR_j}{R^3}\right),\,\,\,K_{ijk}(\boldsymbol{R})=
   \frac{3}{4\pi}\frac{R_iR_jR_k}{R^5}.$$ Using the expression for
 the Helfrich force, we can find from ($\ref{ie}$) the velocity field on the surface of the vesicle.
 The Helfrich force is given in terms of the shape as  \begin{equation}\label{Hf}\begin{split}F_i=&-2
 \left(\kappa[2H(H^2-K)+\Delta_SH]-ZH\right)N_i+\\&+\frac{\partial Z(\boldsymbol{x})}
 {\partial R_j(\boldsymbol{x})}\left(\delta_{ij}-N_iN_j\right),\end{split}\end{equation}
 where $K$ is the Gaussian curvature and $\Delta_S$ is the Laplace-Beltrami operator. It can easily be checked that  expression (\ref{Hf}) vanishes for a spherical shape,  and thus is of order $O(\varepsilon).$ In order to balance (\ref{ie}) at order $O(1)$ we need to assume the imposed flow to be of the same order as (\ref{Hf}). This requirement can equivalently be fulfilled, following Refs. \cite{M06,D07},  by the demand that $\kappa$ scale as $\varepsilon^{-1}.$ We then set $\kappa=\varepsilon^{-1}\bar{\kappa},$ with $\bar{\kappa}$ of order $O(1).$

The precise technical details will be given elsewhere, while here we
shall present only the spirit and some intermediate steps. We expand
$\boldsymbol{v}(\boldsymbol{x}),$ $\boldsymbol{u}(\boldsymbol{x}),$
and $\boldsymbol{F}(\boldsymbol{x})$ into vector spherical harmonics
of $\boldsymbol{x}.$ For convenience we define them as
$$Y^i_{1,k,l}(\boldsymbol{x})=e_{imn}x_m\partial_nY_{k,l}(\boldsymbol{x}),$$
$$Y^i_{2,k,l}(\boldsymbol{x})=(2k+1)x_iY_{k,l}(\boldsymbol{x})-\partial_iY_{k,l}(\boldsymbol{x}),$$
$$Y^i_{3,k,l}(\boldsymbol{x})=\partial_iY_{k,l}(\boldsymbol{x}).$$
Here the differentiation with respect to  $x_i$ is taken formally as
if $\boldsymbol{x}$ were a regular 3D vector not bound to a sphere
of radius unity, and $e_{imn}$ is the Levi-Civita (unit
anti-symmetric) tensor. The advantage of such a definition is that
not only do these spherical harmonics constitute an orthogonal set
with respect to the integration of dot product over
$\boldsymbol{x},$ i.e. the quantity $$\int
Y^i_{j,k,l}(\boldsymbol{x})Y^{i}_{j',k',l'}(\boldsymbol{x})^*d^2x$$
is zero unless $j=j',$ $k=k',$ $l=l',$ but also operators $G$ and
$K$ are diagonal in the chosen basis for a spherical vesicle, so
that the integrals
$$\int\int{Y^i_{j,k,l}(\boldsymbol{x})G_{im}(\boldsymbol{x}-\boldsymbol{x}')Y^{m}_{j',k',l'}(\boldsymbol{x}')^*d^2xd^2x'},$$
$$\int\int{Y^i_{j,k,l}(\boldsymbol{x})K_{imn}(\boldsymbol{x}-\boldsymbol{x}')Y^{m}_{j',k',l'}(\boldsymbol{x}')^*x_nd^2xd^2x'}$$
are zero unless $j=j',$ $k=k',$ $l=l'.$
 Given the centro-symmetry we impose, in our expansion of  $\boldsymbol{v}(\boldsymbol{x}),$ $\boldsymbol{u}(\boldsymbol{x}),$ and $\boldsymbol{F}(\boldsymbol{x}),$  the absence of  $Y^i_{1,k,l}(\boldsymbol{x}),$ for even $k,$ and $Y^i_{2,k,l}(\boldsymbol{x}),$ and $Y^i_{3,k,l}(\boldsymbol{x}),$ for odd $k.$   In order to find the evolution equations we need to expand the velocity field to the order of $O(\varepsilon^2).$ Coefficients for $Y^i_{j,k,l}(\boldsymbol{x})$ for $k>6$ are $O(\varepsilon^3)$ and thus can be neglected. It will be shown later that coefficients of 5th and 6th orders of vector spherical harmonics make corrections to the evolution equations for $f_{k,l}$ with $k\le4$ which are $O(\varepsilon^3).$ So only five first orders
 (starting from the zeroth one) of vector spherical harmonics should be taken into account in the  expansion of any space-dependent quantity under consideration.

In order to fulfil the condition of local inextensibility of the membrane we impose  zero surface divergence of the velocity field\cite{M06}. \begin{equation}\label{sd}\frac{dA(\boldsymbol{x})}{dt}=\frac{\partial v_i(\boldsymbol{x})}{\partial R_j(\boldsymbol{x})}\left[\delta_{ij}-N_i(\boldsymbol{x})N_j(\boldsymbol{x})\right]dA(\boldsymbol{x})=0.\end{equation} Here $$\frac{\partial v_i(\boldsymbol{x})}{\partial R_j(\boldsymbol{x})}=\frac{\partial v_i(\boldsymbol{x})}{\partial x_k}\frac{\partial x_k}{\partial R_j(\boldsymbol{x})}$$ is the Jacobian matrix. We can take the derivatives as if $\boldsymbol{x}$ were a regular 3D vector, because the surface divergence is fully determined by the distribution of the velocity field on the membrane and does not depend on the continuation of the flow into the liquids.

We expand equations (\ref{ie},\ref{sd}) to  order $O(\varepsilon^2)$ and the resulting integrals could be
performed analytically.
Projecting the results of the integration on the space of vector spherical harmonics up to the fourth order, we obtain a set of equations satisfied by the coefficients entering the  expansions of $v(\boldsymbol{x})$ and $Z(\boldsymbol{x}).$ {Since   the surface area is evaluated up to  order  $O(\varepsilon^3),$ while  the velocity field in (\ref{sd}) is expanded only up to  order $O(\varepsilon^2),$ it is not appropriate to use (\ref{sd}) in order to ensure the conservation of the whole surface area. Therefore, we only use the projections of (\ref{sd}) on $Y_{k,l}(\boldsymbol{x})$ for $k$ equal to $2$ or $4.$
Accordingly, we leave the isotropic part of $Z(\boldsymbol{x})$ (denoted here as $Z_0$) undetermined. Once the final shape evolution equations are obtained  we shall  use the constraint that the time derivative of (\ref{ea}) is equal to zero in order to determine  $Z_0.$ This way of reasoning was used in    previous studies \cite{M06,D07,G07,L07}.}
 Note that unlike  other theories \cite{M06,D07,G07,L07}, since we expand the equations to higher order, it is not legitimate  to replace $dA(\boldsymbol{x})$  in (\ref{sd}) with $d^2x$ prior to projection  on the spherical harmonics sub-space. Such a substitution would imply neglecting  terms of order $O(\varepsilon^2)$ in final equations, and would be inconsistent with the spirit of the present theory.

Having determined the velocity field thanks to the above expansions, we are in a position  to obtain the final evolution equation by making use of the  kinematic equation expressing the fact that the membrane velocity is equal to the fluid velocity at the membrane\begin{equation}\label{ke}\frac{\partial f(\boldsymbol{x})}{\partial t}=v_i(\boldsymbol{x})x_i-\frac{\partial_if(\boldsymbol{x})v_j(\boldsymbol{x})(\delta_{ij}-x_ix_j)}{R(\boldsymbol{x})}\end{equation} Then the task is to substitute the expanded velocity field into this equation (in terms of coefficients of the velocity field) and project the resulting expression onto the space of spherical harmonics of interest.
This then leads to the determination of the  evolution equations that must be satisfied by the shape coefficients $f_{k,l}.$ It can be shown that   the following identities hold $Y^i_{1,5,l}(\boldsymbol{x})x_i=0,$ $Y^i_{2,6,l}(\boldsymbol{x})x_i=7Y_{6,l}(\boldsymbol{x}),$ and $Y^i_{2,6,l}(\boldsymbol{x})x_i=6Y_{6,l}(\boldsymbol{x}).$ Therefore   the projection of the first part of the right hand side of (\ref{ke}) on the subspace of spherical harmonics of orders up to four does not depend on the coefficients of the vector spherical harmonics of orders five  and  six in the velocity field. Regarding the  the second part, $\partial_if(\boldsymbol{x})$ is of order $O(\varepsilon)$ and the coefficients for fifth  and sixth orders of vector spherical harmonics in the velocity field expansion are of order $O(\varepsilon^2)$ so that  their contribution in  the shape evolution equation is of order  $O(\varepsilon^3),$ that is beyond our accuracy.

It is convenient to decompose the applied shear flow into its elongational and rotational parts to simplify the final equations. The same decomposition can be applied to general linear flow:
the quantity $E_2(\boldsymbol{x})=U_i(\boldsymbol{x})x_i/2$ defines the straining part, while the vector $\Omega_i=e_{ijk}\partial_jU_k(\boldsymbol{x})/2$ represents the vorticity. Note, that $u(\boldsymbol{x})$ in (\ref{ie}) is the velocity of the imposed flow evaluated at the point $\boldsymbol{R}(\boldsymbol{x}),$ which takes the following form $$u_l(\boldsymbol{x})=(\partial_lE_2+e_{ljk}x_j\Omega_k)(1+f(\boldsymbol{x}))$$

The shape evolution equations for the second and fourth order harmonics can be written in a compact form:
\begin{equation}\label{dotf2}\varepsilon\frac{Df_{2,l}}{Dt}=\frac{\int F_2(\boldsymbol{x})Y_{2,l}(\boldsymbol{x})^*d^2x}{\int Y_{2,l}(\boldsymbol{x})Y_{2,l}(\boldsymbol{x})^*d^2x},\end{equation}
\begin{equation}\label{dotf4}\varepsilon^2\frac{Df_{4,l}}{Dt}=\frac{\int F_4(\boldsymbol{x})Y_{4,l}(\boldsymbol{x})^*d^2x}{\int Y_{4,l}(\boldsymbol{x})Y_{4,l}(\boldsymbol{x})^*d^2x},\end{equation}

The left  hand side term  of the equation is a special  derivative
of $f_{k,l}$ that naturally arises in a non-rotating coordinate
system, and its definition is
\begin{equation}\frac{Df_{k,l}}{Dt}=\frac{\partial f_{k,l}}{\partial
t}+\frac{\int
e_{ijm}\partial_if_kx_j\omega_mY_{k,l}(\boldsymbol{x})^*d^2x}{\int
Y_{k,l}(\boldsymbol{x})Y_{k,l}(\boldsymbol{x})^*d^2x}
\label{jaumann}.\end{equation} To the order at which our expansion
is performed, the rotational velocity of the vesicle is not equal to
the vorticity of the imposed flow (unlike lower order calculations
\cite{M06,G07,D07,L07}), but rather it has a new contribution
originating from the shape function
\begin{equation}\label{omega}\omega_j=\Omega_j+\varepsilon\frac{e_{jkl}\partial_{ik}E_2
\partial_{il}f_2}{2}+O(\varepsilon^2).\end{equation}
Note that if only the first term $\Omega_i$ is retained
\cite{M06,G07,D07,L07}, then (\ref{jaumann}) coincides with the
Jaumann derivative.

The functions $F_k$ can be written as $$F_2=a_1f_2+\varepsilon
a_2f_2^2+\varepsilon^2(a_3f_2^3+a_4f_2f_4)+b_1E_2+\varepsilon
b_2E_2f_2+$$
$$+\varepsilon^2(b_3f_2\partial_if_2\partial_iE_2+b_4\partial_if_2\partial_{ij}f_2\partial_jE_2+$$ $$+b_5\partial_if_2\partial_{ij}E_2\partial_jf_2+b_6E_2f_4)+O(\varepsilon^3),$$
$$F_4=\varepsilon(c_1f_4+c_2f_2^2)+\varepsilon^2(c_3f_2^3+c_4f_2f_4)+\varepsilon d_1E_2f_2+$$ $$+\varepsilon^2(d_2f_4E_2+d_3f_2^2E_2+d_5f_2\partial_iE_2\partial_if_2)+O(\varepsilon^3).$$
 The coefficients $a_i,$ $b_i,$ $c_i,$ and $d_i$ are rational functions of $\lambda,$ $\bar{\kappa}=\varepsilon/C_a,$ and $Z_0.$ The exact expressions are listed in the appendix.

\section{The phase diagram}
\subsection{The analytical phase diagram and comparison with
previous works}

\begin{figure}
\begin{center}
\includegraphics[angle=-90,width=0.9\columnwidth]{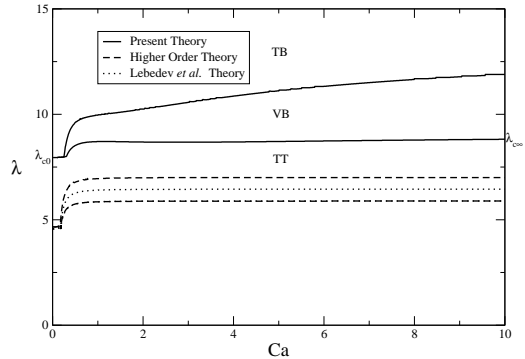}
\caption{\label{figpd} Continuous lines represent the phase diagram
for $\Delta=0.43,$ the results of the higher order theory\cite{D07}
and Lebedev {\it et al.} theory\cite{L07} are added in dashed and dotted lines
for comparison. The TT phase borders are almost indistinguishable for last two theories.}
\end{center}
\end{figure}

The phase diagram for $\Delta=0.43$ is presented on the Fig.1. The
value $\Delta=0.43$ is chosen for the sake of comparison with
available numerical data \cite{B09}. Also shown on that Figure are
the results obtained in two previous theories\cite{D07,L07}. As can
be seen the basic structure of the phase diagram bears similarity
with previous ones. There are however some important distinctions.
First, the phase borders are significantly shifted towards  the
region of higher viscosity contrasts. Second, the VB/TB transition
curve does not saturate upon increasing shear rate (or $C_a$),
rather it exhibits a striking widening. We have also compared the
results with those obtained recently by full three dimensional
simulations. A remarkable agreement between the two approaches is
found (see comparison in Ref. \cite{B09}).

\subsection{Basic reasons for necessity of higher order expansion}

\begin{figure}
\begin{center}
\includegraphics[angle=-90,width=0.9\columnwidth]{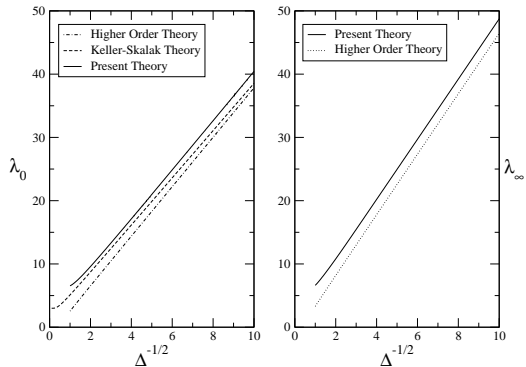}
\caption{\label{figlambdac} $\lambda_0$ and $\lambda_\infty$ as a
function of $\Delta^{-1/2}.$}
\end{center}
\end{figure}

 {Our first concern is to understand why the
inclusion of the terms of order $O(\varepsilon^2)$ provides such a
dramatic shift of the phase borders even though the shape is almost
spherical (the relative excess area is only $0.43/(4\pi)=0.034$). We
have studied the evolution of the phase diagram as a function of
$\Delta$ in order to gain further insight. We have tracked the
viscosity contrast $\lambda$ at which the loss of stability of TT
motion occurs in the two asymptotic limits $C_a\rightarrow 0$ (the
corresponding value is denoted as $\lambda_{c0}$) and $C_a=\infty$
(denoted as $\lambda_{c\infty}$).}
It was reported  in \cite{L07}
that $\lambda_c\propto\varepsilon^{-1}.$ Since we expect the
expansions of $\lambda_c$
 to be analytic in $\varepsilon,$ we have thus attempted the following
ansatz:
\begin{equation}\label{lc}\lambda_c=\lambda_c^{(-1)}\varepsilon^{-1}+\lambda_c^{(0)}+O(\varepsilon).\end{equation}
To check the validity of this expansion  we plot these critical
values as a function of $\Delta^{-1/2}=1/\varepsilon.$ We expect then
a linear behavior. This is presented on Fig.\ref{figlambdac} where
we also  compare our results with those of  previous theories.
 A key
point to be discussed further below is the fact that previous theories provide correct values for the dominant
 term $\lambda_c^{(-1)}$ in the expansions (\ref{lc}), but not for $\lambda_c^{(0)}$ which does not vanish
 even in the limit of almost spherical vesicles (note that the next order terms in the expansion (\ref{lc})
 tend to $0$ with the excess area). Because $\lambda_c^{(0)}$ remains finite whatever small is the deviation
  from a sphere, the discrepancy between the TT phase borders obtained by different theories (with one lower
  order below the present one, namely of the order of $O(\varepsilon)$)
persists for any excess area. In contrast the present theory valid
to order $O(\epsilon^2),$ has the property that even if one wishes
to make an expansion to the next order (i.e. order $O(\epsilon^3)$),
then the shift in the borders in the phase diagram would be
negligibly small, i.e. the shift would vanish in the small excess
area limit. In other words, the present theory shows that there is a
convergence of the small deformation scheme at order $O(\epsilon^2
).$ This implies that  there is no need to continue the expansion of
the shape evolution equation
  beyond the order $O(\varepsilon^2)$ since  this  would neither notably modify the results
  for $\Delta<0.43$ nor significantly extend the applicability to higher values of $\Delta.$
Comparison of the present theory with numerical results (see
comparison provided in Ref. \cite{B09}) shows a good
 agreement for $\Delta=0.43.$ (the only value explored to date by the numerical scheme).

\subsection{Discussion of the tank-treading  phase borders}

As can be seen the dependencies shown on Fig.\ref{figlambdac}  are
fairly linear for $\Delta<1$ and are almost indistinguishable from
straight lines for $\Delta<0.5.$ The region of $1/\sqrt{\Delta}$ where the curves noticeably deviate from straight lines is wider for present theory than for HOT. This is most likely caused by the introduction
of fourth order spherical harmonics, their amplitude is no more
small compared to that of the second order harmonics for $\Delta>1.$
The expression (\ref{ea}) becomes inapplicable, and as a result
$f_{4,l}(t)$ grow uncontrollably with time in solutions of the
differential equations. This problem can be fixed by expanding the
excess area to the order $O(\varepsilon^4),$ but the results of such
patching are not trustworthy because they go beyond the precision of
the differential equations.

Note also that the second term in (\ref{omega}) tends to align the
vesicle to the elongating direction of the straining part of the
shear flow which makes the angle $\pi/4$ with the direction of the
flow. In the TT phase close to the loss of the stability of the
steady-state solution the vesicle is almost parallel to the flow.
The effective vorticity (\ref{omega}) turns out to be less than
$\Omega$ in the TT phase, and it leads to the increased stability of
the steady-state solution. The correction to the vorticity is
proportional to $\varepsilon$ and thus increases with $\epsilon$ and
dominates over $\Omega$ in the large excess area limit. As a
consequence, the truncation (\ref{omega}) becomes illegitimate.
Generally, we may assert that $\Delta<1$ is a good estimate for the
applicability limits of the small deformation approximation.

Some theories\cite{M06,L07,K82} allow for analytical extraction of
the coefficients $\lambda_c^{(-1)}$ and $\lambda_c^{(0)},$ and for
HOT and the present theory the extraction is numerical by using the
 slope and offset of the lines on the Fig.2. The combined data are
presented in the tables 1 and 2.
\begin{table}
\begin{center}
\begin{tabular}{|c|c|c|}\hline Theory&$\lambda_{c0}^{(-1)}$&$\lambda_{c0}^{(0)}$\\\hline
 LOT\cite{M06}&$\frac{8}{23}\sqrt{30\pi} (3.38)$&$-\frac{32}{23}(-1.39)$\\\hline
 HOT\cite{D07}&$3.91$&$-1.22$\\\hline
Lebedev {\it et al.}\cite{L07}&$\frac{16}{23}\sqrt{10\pi}
(3.90)$&$-\frac{32}{23}(-1.39)$\\\hline Keller and Skalak
\cite{K82}&$\frac{2}{3}\sqrt{10\pi}(3.74)$&$\frac{73}{63}(1.17)$\\\hline
Present&3.89&1.54\\\hline\end{tabular}
\caption{$\lambda_{c0}^{(-1)}$ and $\lambda_{c0}^{(0)}$ extracted
from different theories. It should be noted that Lebedev et
al.\cite{L07} theory provided only values for $\lambda_{c0}^{(-1)}$
and we have formally  extracted from their theory the value of
$\lambda_{c0}^{(0)}$ for comparison purpose.}
\end{center}
\end{table}

\begin{table}
\begin{center}
\begin{tabular}{|c|c|c|}\hline Theory&$\lambda_{c\infty}^{(-1)}$&$\lambda_{c\infty}^{(0)}$\\\hline
 LOT\cite{M06}&$\frac{8}{23}\sqrt{30\pi} (3.38)$&$-\frac{32}{23}(-1.39)$\\\hline
 HOT\cite{D07}&$4.79$&$-1.38$\\\hline
Lebedev {\it et al.}\cite{L07}&$\frac{16}{23}\sqrt{15\pi}
(4.78)$&$-\frac{32}{23}(-1.39)$\\\hline
Present&4.76&1.13\\\hline\end{tabular}
\caption{$\lambda_{c\infty}^{(-1)}$ and $\lambda_{c\infty}^{(0)}$
extracted from different theories. It should be noted that Lebedev
et al.\cite{L07} theory provided only values for
$\lambda_{c\infty}^{(-1)}$ and we have formally  extracted from
their theory the value of $\lambda_{c\infty}^{(0)}$ for comparison
purpose.}
\end{center}
\end{table}

HOT and Lebedev {\it et al.} theories show a good agreement for the
coefficients $\lambda_{c}^{(-1)}$ with the present calculation. This
means, that expansion of the shape evolution equations to the order
$O(\varepsilon)$ was indeed sufficient in order to capture the
correct value of  $\lambda_{c}^{(-1)}$ even for large values of
$C_a,$ where Lebedev {\it et al.} theory loses its applicability.
The present theory  shows that the correct value of
$\lambda_{c}^{(0)}$ is captured thanks to the fact that the
expansion scheme is pushed one step further in the excess area.
Keller-Skalak theory gives a slightly different result for
$\lambda_{c}^{(-1)},$ the difference is caused by the fact they used
a  velocity field having a surface divergence of order
$O(\varepsilon),$ whereas the membrane local inextensibility
requires zero surface divergence instead. On the other hand
Keller-Skalak theory provides a better estimate for the value of
$\lambda_{c}^{(0)}$ than the one proposed by HOT and this explains
why its results are relatively close to the results of numerical
simulations for $\Delta\sim 1.$

\subsection{Discussion of TB/VB transition} We plan to dedicate a separate research to
the properties of different types of vesicle motion, so here we only
briefly discuss further implication of the new theory on the  TB/VB
transition. Neglecting harmonics of orders higher than two, as done
in previous theories,  provides two major simplifications. First,
the vesicle shape has three symmetry planes, which make the
definition of the orientation angle obvious, and second the in-plane
motion is defined by
 only two independent variables. As a consequence,
the dynamics relaxes with time to a cyclic motion (which degenerates
into a point in the TT phase). Thus only a simple limit cycle can
exist, and this shows that each of the VB and TB modes has its own
region of existence in parameter space. The above  assertions can
not be made  in general, and especially when the fourth order
harmonics are included.
It also turns out that near the transition region the vesicle tends
to finish its TB quasi-cycles by assuming an oblate almost
axisymmetric shape in the shear plane rather than an elongated shape
perpendicular to the flow. For such shape the error in the
definition of the orientation angle is very large.

Let us now discuss how each type of motion is determined from our
evolution equations. We start with some initial values for $f_{kl}
(t)$ and wait a ceratin time interval until  the initial data are
irrelevant. Then it was checked whether during one oscillation
quasi-cycle $f(\boldsymbol{x})$ has a maximum for $\boldsymbol{x}$
lying in the shear plane and perpendicular to the shear velocity.
Despite some noise due to the aforementioned complications it is
clearly seen that unlike with previous theories the VB/TB phase
border does not saturate even for $C_a=10,$ and the VB region
broadens with the increase of the capillary number (Fig.
\ref{figpd}).

Note that it was suggested recently \cite{V09} that  higher orders
of spherical harmonics can be excited close to the VB/TB phase
border and this  may cause some widening of  the VB phase region
(but still saturation at large $C_a$ is found, unlike our theory).
It can be checked that when  $Z_0+20\epsilon/C_a <0$,  the fourth
order harmonics are excited (i.e. the corresponding decay time
becomes infinite to leading order). Here, we found that close to the
VB/TB transition at some times during the oscillation the quantity
\begin{equation}\label{z0c1}-\frac{Z_0C_a}{\varepsilon}\end{equation}
exceeds 20. This is probably  the
reason why the VB/TB phase border remains unsaturated for much
larger values than in previous analytical studies. It was proposed
in \cite{V09}, however, that  excitation of higher harmonics was due
to thermal fluctuations. This contrasts in spirit with our theory
where  the fourth order harmonic is excited as non-linear
interaction of second harmonics and no reference to  temperature is
required.

{
\section{Comparison with experiments}
Experiments have been performed recently regarding determination of the phase diagram of vesicle motions under shear flow\cite{De09}. They referred to a previous theoretical work \cite{L07} which had suggested that the phase diagram should depend on only two independent dimensionless control parameters (and not three), namely   \begin{equation}\label{L}\Lambda=\frac{4}{\sqrt{30\pi}}\left(1+\frac{23}{32}\lambda\right)\sqrt{\Delta},\,\,S=\frac{7\pi}{3\sqrt{3}}
\frac{C_a}{\Delta}\end{equation} It has been reported \cite{De09} (with a certain degree of uncertainty) that the experimental
data were consistent with the fact that only the above two parameters determine the phase diagram.  Our results do not comply with this report, as shown on Figure \ref{phase_diag_Delta}. Indeed, besides $S$ and $\Lambda$, the excess area $\Delta$ plays an important role. Experiments mixed  data for different $\Delta$'s in the plane $(S,\Lambda)$. The band of the VB mode in experiments looks quite wide, and we believe that this reflects the sensitivity of the location of the VB band to excess area (in other words the experimental data may be viewed as  juxtaposition of bands each representing a value of $\Delta$; see the example of    Figure \ref{phase_diag_Delta}).
 It is hoped that
a study representing each value of $\Delta$ will be performed in the future with the aim of making  comparison with theory clearer.

\begin{figure}
\begin{center}
\includegraphics[angle=-90,width=0.9\columnwidth]{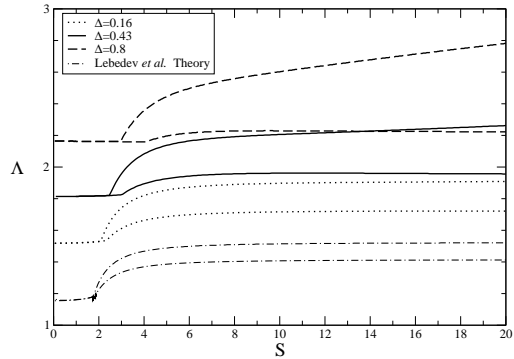}
\caption{\label{phase_diag_Delta} Phase diagrams for various values of $\Delta$ in $S-\Lambda$ coordinates.}
\end{center}
\end{figure}

Furthermore, it must be noted  that our estimates for the TT-VB phase borders are higher than those  observed in experiments. This is not very surprising since transients are found to be very long close to TT-VB transition (as  also discussed in our recent full numerical simulations \cite{B09}), and therefore a firm conclusion on the type of motion can not be made on the basis of the current experimental data, which are limited to only few periods of oscillation (while transient can exhibit up to hundreds of cycles as discussed recently \cite{B09}).  In other words,  this kind of long relaxation would convey the impression that dynamics is of VB type, whereas in reality it is a TT one.



\begin{figure}
\begin{center}
\includegraphics[angle=-90,width=0.9\columnwidth]{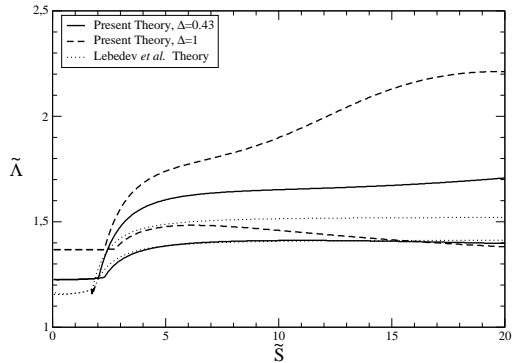}
\caption{\label{other_phase} Phase diagrams of vesicle motions under general planar linear flows. The viscosity contrast $\lambda$ is fixed to $1$ and the ratio of rotating and straining parts of the flow $\omega_0/s_0$ is varied. The theory of Lebedev {\it et al.}\cite{L07} is plotted in dotted lines for comparison.}
\end{center}
\end{figure}

\begin{figure}
\begin{center}
\includegraphics[angle=-90,width=0.9\columnwidth]{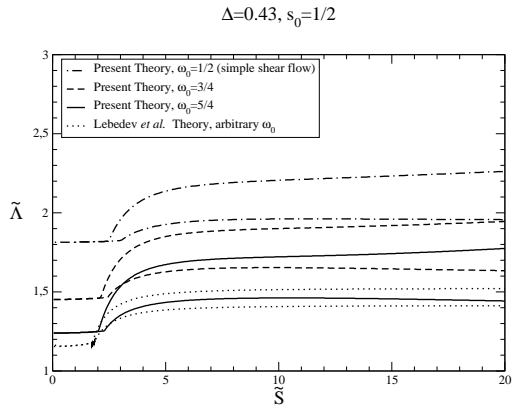}
\caption{\label{phase_diag_general} Phase diagrams of vesicles under general planar linear flows. The straining and rotational parts of the flow are denoted as $s_0$ and $\omega_0$ respectively.}
\end{center}
\end{figure}

Other  experiments\cite{De092} were performed  under linear flow with applied velocity field $u_x=(s_0+\omega_0)y,$ $u_y=(s_0-\omega_0)x,$ $u_z=0$, which is different from a simple shear flow. The experiments were performed with no viscosity contrast and the ratio of the rotational ($\omega_0$) and the straining ($s_0$) parts of the flow
was varied instead. The results of the experiments were plotted in coordinates
\begin{equation}\label{L1}\tilde{\Lambda}=\frac{4\omega_0}{\sqrt{30\pi}s_0}\left(1+\frac{23}{32}\lambda\right)\sqrt{\Delta},\,\,\tilde{S}=\frac{14\pi}{3\sqrt{3}}
\frac{s_0}{\kappa\Delta}\end{equation} representing the
generalization of (\ref{L}). It was then claimed \cite{De092} that
the phase diagram depends on these two parameters only (at least
with no viscosity contrast). The Fig.\ref{other_phase} simulates
this experiment using the present theory. The discrepancy between
resulting phase diagrams for various values of $\Delta$ is indeed
not as striking as for the simple shear flow
(Fig.\ref{phase_diag_Delta}): the TT phase borders fall close to
each other and VB bands overlap for a wide range of $\Delta.$ At the
same time, the VB/TB phase borders still vary significantly with the
excess area. However, by fixing  $\Delta$ and varying  $\lambda$ we
can produce series of phases diagrams in the $\tilde S$ and $\tilde
\Lambda$ plane corresponding to different values of $\omega_0/s_0,$
(see Fig.\ref{phase_diag_general}). Here we find  the same kind of
discrepancy as on Fig.\ref{phase_diag_Delta} .

Finally, let us compare  our results with other set of  experiments in the TT regime\cite{K06}. We represent  the vesicle orientation angle $\phi_0$ under strong shear flow(Fig. 3) (for the sake of comparison with experiments which were performed at high shear rates). It can be checked that the exact form of the bending energy becomes insignificant under strong flows, because the dominant contribution to the membrane force comes form the tension part which enforces incompressibility of the membrane. Unlike for the VB/TB phase border, the dependence of $\phi_0$ on $\lambda$ quickly  saturates  with $C_a.$ We take $C_a=100$ in our calculation. We use the artificial parameter $\Lambda$ (\ref{L}) instead of $\lambda$ in order to show the effect of $O(\epsilon^2)$ terms taken into account by the present calculation. Note that the results of Lebedev {it et al.} theory \cite{L07} do not depend on the excess area $\Delta$ in this representation, unlike our results and those reported by experiments (see Fig. 5). The results of HOT depend also on $\Delta$, but that dependence is weak so that they hardly differ from theoretical results of Ref.\cite{L07} (we do not plot them here in order
 to avoid encumbering of the figure). Experimental results\cite{K06} are provided  for four different values of $\Delta:$ $0.15,$ $0.24,$ $0.42,$ and $1.43.$ We exclude $\Delta=1.43$, since we do not expect our theory to be applicable to a such large value of excess area.  As can be seen,  theoretical results   and experimental ones show a rather good agreement  provided that one is not close to the TT/VB border (i.e. if the angle is not too close to zero).  The discrepancy for small inclination angles may  be attributed  to thermal fluctuations or interactions with walls. However,  a systematic theoretical  study of these factors should be performed before drawing conclusive answers.

}


\begin{figure}
\begin{center}
\includegraphics[angle=-90,width=0.9\columnwidth]{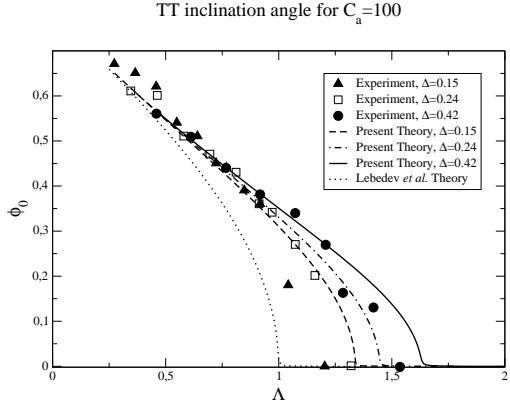}
\caption{\label{figangle} inclination angle of vesicles in TT phase for large $C_a.$}
\end{center}
\end{figure}

\section{Discussion and conclusion}
Here we have presented a small deformation theory of a vesicle in shear flow keeping one more order
in the expansion of the shape evolution equations than prior studies.
  We have confirmed the leading term in the asymptotic expansion of
 critical viscosity ratios determined by previous analytical studies, but we also were able to
 determine accurately the next term in the expansion. This term is constant and survives no matter how small
  the excess area is. Moreover, this theory   provides a correction to the phase borders that is significant
   in a wide range of excess areas.
 Unlike  previous analytical works, but in agreement with
 the results of the numerical simulations \cite{B09}, we observe an unsaturated  growth
 of the VB/TB phase border in a
 quite large  range of the capillary number.

We hope that this work will incite future experimental research.  It will
also be interesting to investigate experimentally the presence of
widening of the VB band as a function of shear rate.  Finally, we
did not include third order spherical harmonics in the expansion of
the shape function, restricting our calculations only to
centro-symmetric vesicles. In the full numerical simulation
\cite{B09}, no symmetry restriction is imposed and so far no
manifestation of the third order harmonic has been observed. This
confers to our assumption of centro-symmetry a certain legitimacy.

\noindent {\bf Acknowledgements:}
We would like to thank G.\,Danker, S.\,S.\,Vergeles, and P.\,M.\,Vlahovska for helpful discussions. A.F. and C.M. acknowledge financial support from CNES (Centre National D'études Spatiales) and ANR (Agence Nationale pour la Recherche); "MOSICOB project".

\appendix
\section{The shape evolution equations and the expression of various coefficients in terms of physical parameters}
We remind that $$F_2=a_1f_2+\varepsilon a_2f_2^2+\varepsilon^2(a_3f_2^3+a_4f_2f_4)+b_1E_2+\varepsilon b_2E_2f_2+$$ $$+\varepsilon^2(b_3f_2\partial_if_2\partial_iE_2+b_4\partial_if_2\partial_{ij}f_2\partial_jE_2+$$ $$+b_5\partial_if_2\partial_{ij}E_2\partial_jf_2+b_6E_2f_4)+O(\varepsilon^3),$$ $$F_4=\varepsilon(c_1f_4+c_2f_2^2)+\varepsilon^2(c_3f_2^3+c_4f_2f_4)+$$ $$+\varepsilon d_1E_2f_2+\varepsilon^2(d_2f_4E_2+d_3f_2^2E_2+d_5f_2\partial_iE_2\partial_if_2)+O(\varepsilon^3).$$
The coefficients are then written as
$$a_1=-24\frac{Z_0+6\bar{\kappa}}{23\lambda+32}$$
$$a_2=24\frac{(49\lambda+136)Z_0+(432\lambda+1008)\bar{\kappa}}{(23\lambda+32)^2}$$
$$b_1=\frac{120}{23\lambda+32}$$
$$b_2=2400\frac{\lambda-2}{(23\lambda+32)^2}$$
$$b_6=-40\frac{241\lambda+344}{(23\lambda+32)^2}$$
$$c_1=-40\frac{Z_0+20\bar{\kappa}}{19\lambda+20}$$
$$c_2=\frac{16}{3}\frac{(92-3\lambda)Z_0+(1822\lambda+3112)\bar{\kappa}}{(19\lambda+20)(23\lambda+32)}$$
$$d_1=\frac{20}{3}\frac{1047\lambda+1072}{(23\lambda+32)(19\lambda+20)}$$
$$d_2=10\frac{257\lambda-32}{(23\lambda+32)(19\lambda+20)}$$
$$d_3=\frac{1}{6}\frac{-125055\lambda^3+594716\lambda^2+1107168\lambda+392320}{(2\lambda+5)(23\lambda+32)(19\lambda+20)^2}$$
\begin{widetext}
$$a_3=-\frac{8}{175}\frac{(2479595 \lambda^3 + 6703156 \lambda^2 + 18601472 \lambda + 16622592}{(23\lambda+32)^3(19\lambda+20)}Z_0-$$
$$-\frac{8}{175}\frac{78181390 \lambda^3  + 390845256 \lambda^2 + 713624832 \lambda + 429094912}{(23\lambda+32)^3(19\lambda+20)}\bar{\kappa}$$
$$a_4=4\frac{(18335 \lambda^2  + 57376 \lambda + 44224)Z_0+(318896 \lambda^2 + 1026064 \lambda + 809600 )\bar{\kappa}}{(19\lambda+20)(23\lambda+32)^2}$$
$$b_3=\frac{1}{70}\frac{78420885 \lambda^4 + 815632786 \lambda^3 + 2193572112 \lambda^2 + 1954954752 \lambda + 481607680 }{(2\lambda+5)(19\lambda+20)(23\lambda+32)^3}$$
$$b_4=-\frac{1}{630}\frac{490295475 \lambda^4  + 3878418742 \lambda^3  + 9801602064 \lambda^2  + 9403966464 \lambda+2954997760 }{(2\lambda+5)(19\lambda+20)(23\lambda+32)^3}$$
$$b_5=\frac{1}{315}\frac{136559745 \lambda^4  + 385825454 \lambda^3  + 141580368 \lambda^2  - 26564352 \lambda + 98232320}{(2\lambda+5)(19\lambda+20)(23\lambda+32)^3}$$
$$c_3=-\frac{4}{63}\frac{2552583\lambda^3+1777744\lambda^2-913584\lambda+315392}{(23\lambda+32)^2(19\lambda+20)^2}Z_0-$$
$$-\frac{16}{63}\frac{32203871\lambda^3+118114286\lambda^2+154157080\lambda+69803008}{(23\lambda+32)^2(19\lambda+20)^2}\bar{\kappa}$$
$$c_4=2\frac{(21383\lambda^2+63844\lambda+44928)Z_0+48(13411\lambda^2+35444\lambda+23040)\bar{\kappa}}{(23\lambda+32)(19\lambda+20)^2}$$
$$d_4=\frac{1}{42}\frac{41587815\lambda^4+95846332\lambda^3-88522016\lambda^2-295841536\lambda-152842240}{(2\lambda+5)(23\lambda+32)^2(19\lambda+20)^2}$$
\end{widetext}

\end{document}